\documentclass[twocolumn, secnumarabic, amsmath, amssymb, nobibnotes, aps, pra]{revtex4-2}

\usepackage{graphicx}
\usepackage{dcolumn}
\usepackage{bm}

\usepackage{upgreek}
\usepackage{amsmath}
\usepackage{braket}
\usepackage{xcolor}
\usepackage{ulem}
\usepackage{soul}
\usepackage[colorlinks=true, linkcolor=blue, urlcolor=blue]{hyperref}

\begin{document}

\preprint{}

\title{Reconfigurable measurement scheme compatible with MDI-QKD and BB84 protocols using single-laser sideband generation}

\author{Jaesung Lim}
\author{Yonggi Jo}
\author{Nam Hun Park}
\author{Zaeill Kim}
\author{Yong Sup Ihn}
\email{yong0862@add.re.kr}
\affiliation{Agency for Defense Development, Daejeon 34186, Korea}
\date{\today}

\begin{abstract}
A quantum network can be operated efficiently by switching between measurement-device-independent quantum key distribution (MDI-QKD) and BB84 according to the trust level of an intermediate node, since MDI-QKD is immune to all detector side-channel attacks while generally yielding a lower secret-key rate.
Here, we present a proof-of-principle demonstration of a reconfigurable measurement scheme compatible with both protocols using a single laser.
Two mutually independent weak coherent states (WCSs) are generated from a shared continuous-wave laser via electro-optic phase modulation and subsequent first-order sideband filtering.
Channel indistinguishability is verified using two complementary Hong--Ou--Mandel (HOM) measurements: time-resolved coincidence measurements and polarization mismatch scans, both yielding HOM visibilities close to the theoretical limit of 0.5 for WCSs.
The measurement module performs a partial Bell-state measurement for MDI-QKD, while rotating a single half-wave plate to $22.5^{\circ}$ reconfigures the same module into a dual-basis polarization analyzer for BB84.
The quantum bit error rates measured in both configurations agree well with theoretical expectations, verifying the operational feasibility of the reconfigurable receiver at the physical layer.
By eliminating the need for active frequency locking while retaining independent sideband tunability, this approach provides a practical foundation for flexible and cost-effective quantum networks. 
\end{abstract}

\maketitle

\section{Introduction}

When two indistinguishable photons impinge on a beam splitter, bosonic bunching suppresses coincident detection at the two output ports.
This phenomenon, known as the Hong--Ou--Mandel (HOM) dip, arises when two input photons are perfectly matched in all relevant degrees of freedom, including polarization, spectrum, spatial mode, and arrival time~\cite{hong1987measurement, bouchard2021two}.
While initially demonstrated with nonclassical biphoton states from spontaneous parametric down-conversion (SPDC), HOM interference can also be realized using weak coherent states (WCSs)~\cite{legero2003time, kim2013conditions, kim2014multimode, ferreiradasilva2015spectral, moschandreou2018experimental, Kim2020HOM, Kim2021APL}.
Although the Poissonian photon-number statistics of WCSs impose a theoretical visibility upper bound of 50\%, the technological maturity of coherent sources makes WCS-based two-photon interference (TPI) a practical building block for quantum information protocols~\cite{Xu2020RMP}.

A prominent application is measurement-device-independent quantum key distribution (MDI-QKD)~\cite{lo2012mdi}.
While theoretical QKD guarantees unconditional security based on quantum mechanics, real-world physical device imperfections can expose critical side-channel vulnerabilities~\cite{lydersen2010hacking,gerhardt2011fullfield}.
Although device-independent QKD solves this fundamentally, it demands near-unity detection efficiencies that remain challenging for practical deployment~\cite{acin2007device}.
In contrast, MDI-QKD circumvents all detector side channels while maintaining compatibility with realistic experimental constraints~\cite{rubenok2013real,liu2013experimental,tang2014polarization,tang2016network}.
However, MDI-QKD generally yields lower key generation rates compared to conventional decoy-state BB84 protocols \cite{Hwang2003,Ma2005,Zhao2006} because it relies on two-photon coincidences.

This intrinsic trade-off motivates reconfigurable QKD platforms that can switch between MDI-QKD and BB84 according to the level of trust assigned to intermediate nodes~\cite{qi2015freespace, fanyuan2021nonstandalone}.
In this work, we present a proof-of-principle demonstration of a reconfigurable measurement scheme compatible with both MDI-QKD and BB84 protocols using electro-optic modulator (EOM)-based sideband generation from a single laser.
By splitting the output of a single continuous-wave (CW) laser and driving the EOMs with independent radio-frequency (RF) signals, two weak coherent state (WCS) channels with mutually independent phases are prepared.
Because both channels originate from a shared laser, they are intrinsically frequency-matched, which eliminates the need for the complex active frequency-locking systems typically required in conventional MDI-QKD implementations.
Furthermore, EOM-based sideband modulation provides high optical transmission efficiency and independent frequency tunability for each transmitter~\cite{lo2017precise}.
To validate channel indistinguishability, we perform time-resolved Hong-Ou-Mandel (HOM) interference and polarization mismatch scans, obtaining visibilities close to the theoretical limit for WCSs.
At the receiver stage, the measurement module functions as a partial Bell-state measurement (BSM) setup for MDI-QKD, while rotating a half-wave plate (HWP) to $22.5^{\circ}$ transforms it into a polarization analyzer for BB84.
The QBERs measured in both configurations agree well with theoretical expectations, verifying the operational feasibility of the reconfigurable measurement scheme at the physical layer and paving the way toward flexible quantum networks.

\begin{figure*}[t]
  \centering
  \includegraphics[width=1\textwidth]{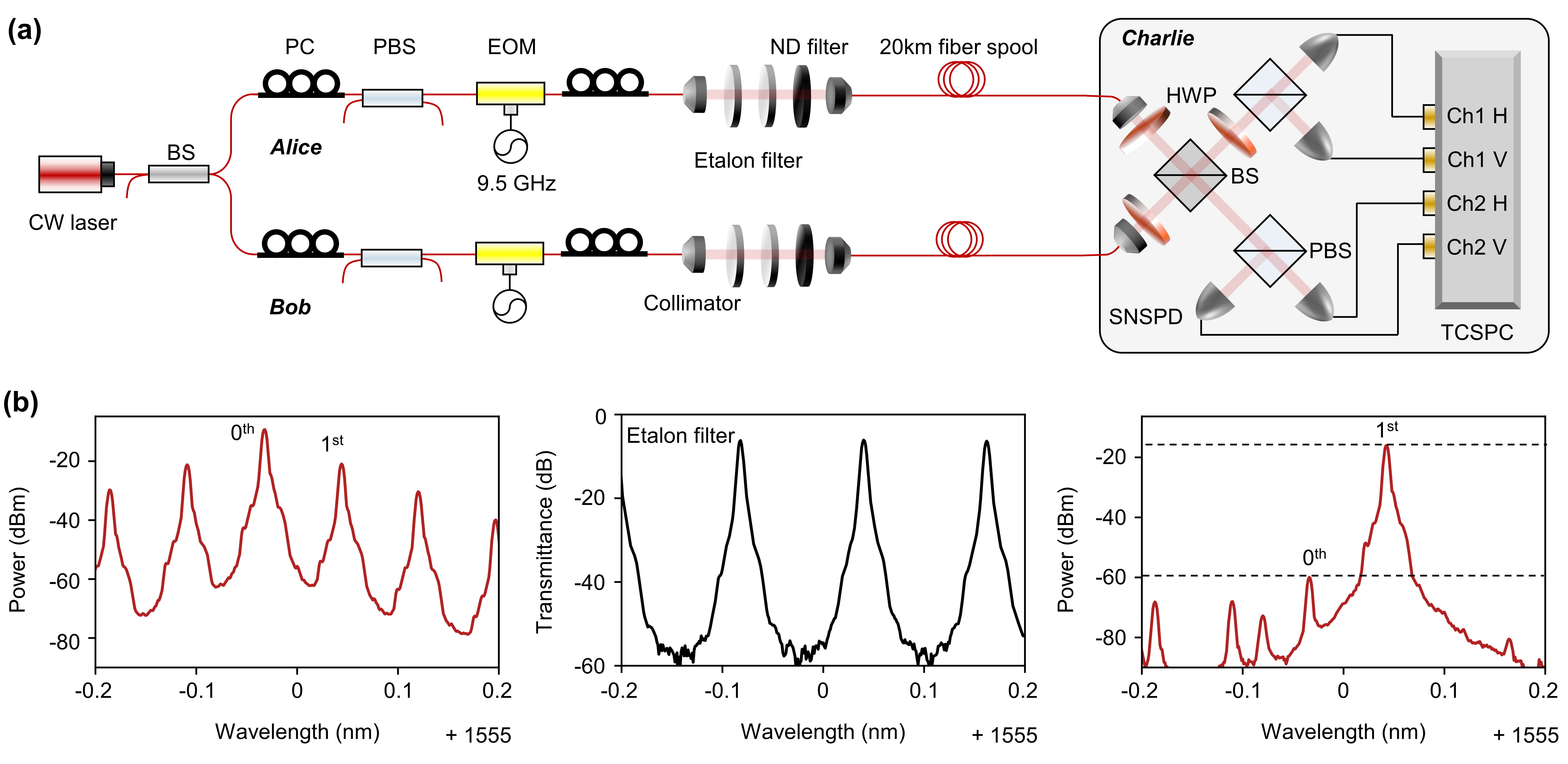}
  \caption{\label{fig1} Experimental setup and spectral characterization of the reconfigurable measurement.
  (a) Schematic of the optical setup.
  Two mutually independent WCSs are generated from a single CW laser using EOMs driven by independent 9.5 GHz RF sources, followed by sideband filtering. After propagating through a 20~km optical fiber spool, the signals enter Charlie's partial BSM module.
  The system is reconfigured simply by toggling an HWP in one arm of the partial BSM module between $0^{\circ}$ and $22.5^{\circ}$.
  (b) The spectra of the WCS preparation.
  The leftmost panel shows the raw optical spectrum after EOM phase modulation. The middle panel displays the transmission profile of two cascaded etalon filters. The rightmost panel confirms the isolation of the first-order sideband, demonstrating residual carrier suppression of approximately 40~dB.}
\end{figure*}

\section{Experimental setup}

\begin{figure*}[t]
  \centering
  \includegraphics[width=1\textwidth]{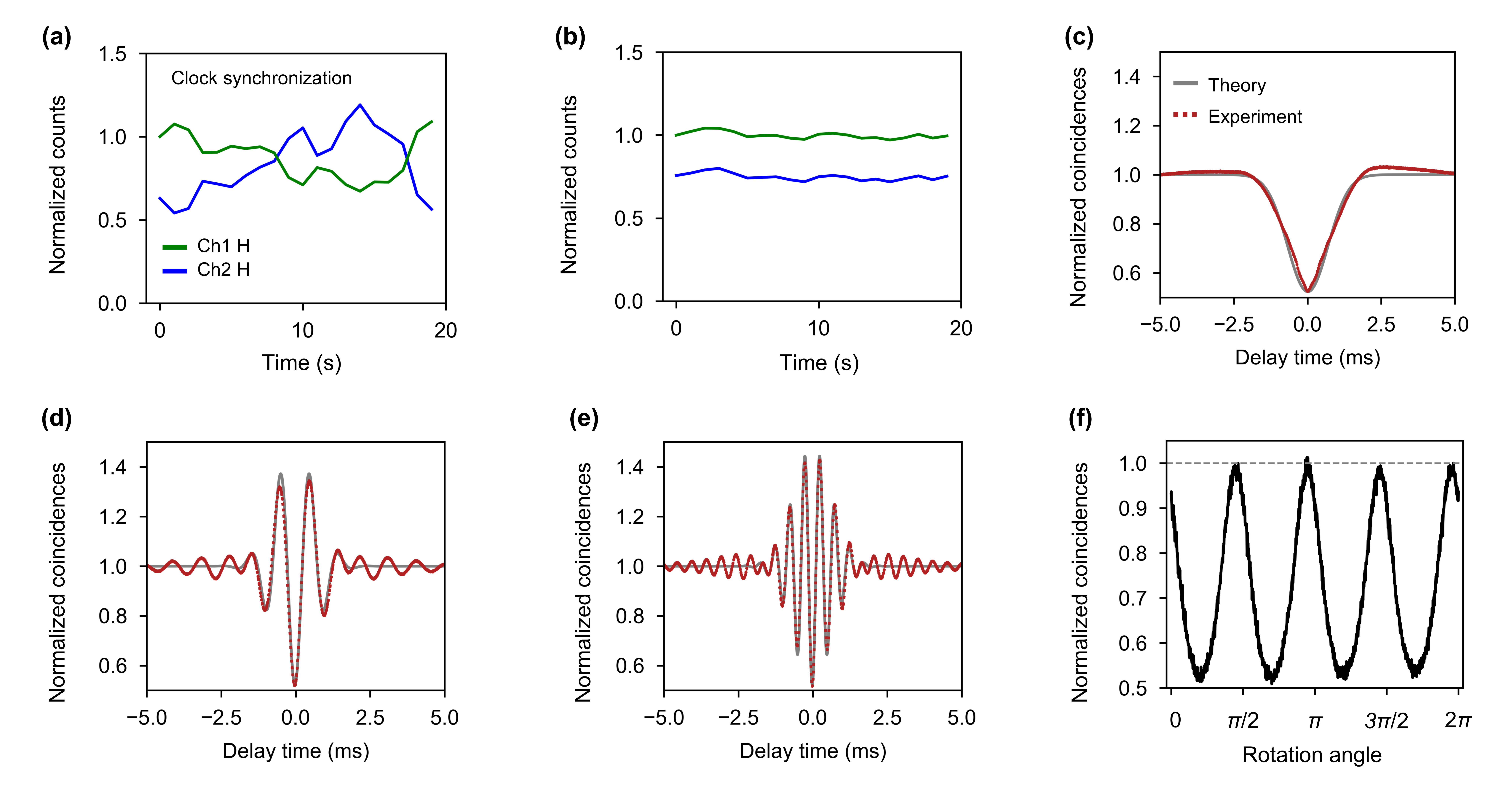}
  \caption{\label{fig2} Experimental verification of mutual independence and two-photon indistinguishability.
  (a,b) First-order interference traces recorded at the H-polarized outputs without the 20~km fiber spool, obtained with (a) synchronized and (b) independent RF source clocks.
  (c--e) Time-resolved HOM coincidence histograms as a function of the detection time difference $\Delta t$ for frequency detunings of (c) 0~kHz, (d) 1~kHz, and (e) 2~kHz in Alice's arm.
  The experimental data (red points) are well described by the theoretical fits (solid gray curves) based on Eq.~(\ref{eq:tpi}).
  (f) A polarization mismatch scan demonstrating a four-period coincidence oscillation. The upper and lower bounds (gray dashed lines) are consistent with the HOM visibility limit.}
\end{figure*}

The schematic of the reconfigurable measurement setup is illustrated in Fig.~\ref{fig1}(a).
A tunable distributed feedback (DFB) laser operating near 1550 nm serves as the common CW light source.
The laser output is split via a 50:50 beam splitter (BS) to form transmitter channels for Alice and Bob.
In each channel, a polarization controller (PC) and a polarizing beam splitter (PBS) are positioned to optimize the input polarization and precisely adjust the optical intensity before the beam enters the EOM.
The EOMs are driven by independent 9.5~GHz sinusoidal RF sources.
Since the intrinsic phase noise of each RF source is completely uncorrelated, the sidebands generated in the two channels become mutually phase-independent.
It should be noted that the optical states prepared here are mutually phase-incoherent WCSs sufficient for two-photon interference, rather than the phase-randomized Poissonian mixtures strictly required for the decoy-state method.
The spectral engineering process is characterized in Fig.~\ref{fig1}(b).
As shown in the leftmost panel of Fig.~\ref{fig1}(b), the raw output from the EOM exhibits a series of sidebands generated at intervals of 9.5~GHz centered around the carrier frequency (0th order).
To isolate a single sideband (1st order), the modulated light is passed through two cascaded etalon filters.
These etalons feature a free spectral range (FSR) of approximately 15 GHz and an extinction ratio close to 50 dB, as detailed in the middle panel of Fig.~\ref{fig1}(b).
The resulting filtered spectrum, shown in the right panel, indicates that the residual carrier is suppressed by approximately 40 dB relative to the selected first-order sideband, confirming effective sideband isolation.
Since both the DFB laser and the EOM drive are tunable, the center wavelength and the sideband spacing can be adjusted independently, allowing WCS channels to be generated in different wavelength bins without any hardware modification.

Following sideband generation and transmission through the 20~km fiber spool, Alice and Bob adjust their PCs to ensure horizontally polarized output.
The four polarization states ($\ket{H}$ and $\ket{V}$ for the Z basis; $\ket{D}$ and $\ket{A}$ for the X basis) are then prepared using motorized HWPs placed immediately before Charlie's measurement node.
To simplify the compensation of fiber-induced polarization drift, this proof-of-principle configuration encodes polarization states after fiber transmission, departing from standard QKD implementations where polarization encoding is performed prior to channel propagation.
The optical signals are attenuated to the weak-coherent regime via neutral density (ND) filters.
Charlie's measurement setup features a partial BSM module capable of identifying the $\ket{\psi^{+}}$ and $\ket{\psi^{-}}$ Bell states.
Incoming photons are detected by four superconducting nanowire single-photon detectors (SNSPDs), with the detection events recorded by a time-correlated single-photon counter (TCSPC) using a 10~ns coincidence window.
Since the laser source operates in CW rather than pulsed mode, the mean photon number per pulse, $\mu$, is not physically well-defined in our setup.
Following established methodologies for CW two-photon interference experiments~\cite{kim2014multimode, Kim2020HOM}, we characterize the two-photon interference and state projections using normalized coincidence count rates.

The reconfiguration is performed by rotating a single HWP, which is inserted in one output arm of Charlie's partial BSM module [Fig.~\ref{fig1}(a)].
When this HWP is set to $0^{\circ}$, the hardware functions as a standard MDI-QKD receiver.
Rotating this HWP to $22.5^{\circ}$ converts the Z-basis measurement in that arm into an X-basis projection.
This architectural shift enables Charlie to measure in both bases simultaneously, effectively implementing the BB84 measurement framework using the same physical hardware~\cite{qi2015freespace, fanyuan2021nonstandalone}.
Furthermore, to avoid contamination of the BB84 measurement by optical signals from the inactive link, the unused input is blocked.

\section{Results and Discussion}

\begin{figure*}[t]
  \centering
  \includegraphics[width=1\textwidth]{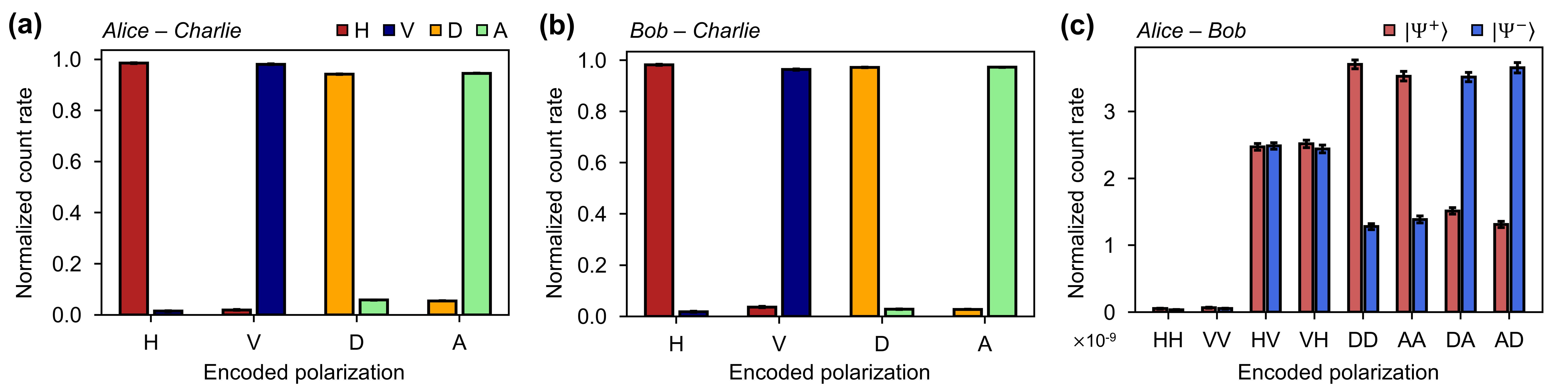}
  \caption{\label{fig3} Measurement results for the MDI-QKD and BB84 configurations. (a,b) Normalized single-photon count rates measured in the BB84 mode, achieved by setting the internal HWP to $22.5^{\circ}$, for the (a) Alice--Charlie and (b) Bob--Charlie links.
  (c) Normalized coincidence count rates for the $\ket{\psi^{+}}$ and $\ket{\psi^{-}}$ Bell state projections in the MDI-QKD mode (HWP at $0^{\circ}$).
  All data points in both operational modes were accumulated for 5~min.}
\end{figure*}

Prior to evaluating TPI, it is critical to verify the mutual independence of the two prepared WCS channels.
We analyzed the first-order interference patterns at the H-polarized outputs of channels 1 and 2 without inserting the 20~km fiber spool.
Photon count traces were measured for 20~s using a bin width of 1~s.
Fig.~\ref{fig2}(a) illustrates the trace obtained when the RF generator clocks are externally synchronized.
In this configuration, the relative phase between the sidebands drifts slowly, resulting in clear first-order interference over time.
In contrast, as depicted in Fig.~\ref{fig2}(b), when the RF generators operate on independent internal clocks, the complete absence of first-order interference patterns demonstrates their mutual independence.

Spatial scanning used in typical TPI experiments cannot characterize photon indistinguishability due to the long mutual coherence time of CW WCSs.
Instead, we performed time-resolved TPI by recording coincidences as a function of the detection time difference $\Delta t$ between the two output ports~\cite{ferreiradasilva2015spectral, Kim2021APL}.
When polarizations and intensities are matched, the normalized coincidence rate is described by
\begin{equation}\label{eq:tpi}
  g^{(2)}(\Delta t) = 1 - V \cos(\Delta\omega\,\Delta t)\,\exp\!\left(-\frac{\Delta t^{2}}{T_{c}^{2}}\right),
\end{equation}
where $V$ represents the visibility, $\Delta\omega$ is the angular frequency difference between the two channels, and $T_{c}$ is the mutual coherence time.

Coincidence histograms were acquired with 1~$\mu$s time bins over a 10~ms temporal window and accumulated for 1~min.
Figs.~\ref{fig2}(c),~\ref{fig2}(d), and~\ref{fig2}(e) show these time-resolved TPI measurements corresponding to frequency detunings of 0, 1, and 2~kHz, respectively.
For zero frequency detuning (Fig.~\ref{fig2}(c)), the minimum normalized coincidence level reaches 0.524, which corresponds to a visibility of $V = 0.476$, approaching the classical limit of 0.5 for WCSs.
Fitting Eq.~(\ref{eq:tpi}) to the data yields an average mutual coherence time of \(T_c = 0.958\)~ms.
This value represents an effective mutual coherence time determined primarily by the phase noise of the independent RF generators and the phase fluctuations accumulated within the 20~km fiber spool.

To further validate channel indistinguishability, we conducted a polarization mismatch scan.
With both channels initially prepared in the $\ket{H}$ state, an HWP in Alice's arm was continuously rotated.
As the angle $\theta$ is swept from 0 to $2\pi$, the polarization returns to $\ket{H}$ four times.
Fig.~\ref{fig2}(f) shows the corresponding coincidence counts.
The expected four-period oscillation is clearly observed, with maximum and minimum boundaries corresponding to the 0.476 visibility obtained from the time-resolved TPI.

We next measure the QBERs of both the MDI-QKD and BB84 configurations as a proof-of-principle demonstration.
We note that the measurements reported here constitute a simplified implementation to verify the feasibility of the reconfigurable receiver at the physical layer, rather than a fully realized QKD system.
For implementation of the full protocol, several modifications are required at the physical layer. First, the system must operate in pulsed mode with individual-pulse phase randomization via an EOM, ensuring a Poissonian mixture of photon-number states for decoy-state security analysis.
In addition, each channel requires a polarization modulator for rapid state preparation and an intensity modulator for decoy-state generation, all synchronized with the pulsed source.
Furthermore, for MDI-QKD operation, the signal pulses from Alice and Bob must arrive simultaneously at the partial BSM with complete indistinguishability in all non-encoded degrees of freedom, which can be achieved by using adjustable optical delay lines.
Integrating these components would provide a clear path toward upgrading the present setup into a complete and secure QKD system.

For the BB84 mode (HWP set to $22.5^{\circ}$), data for each encoded polarization state were accumulated over 5~min.
In this setup, the QBER is evaluated as:
\begin{equation}\label{eq:qber}
  E = \frac{C_{\perp}}{C_{\parallel} + C_{\perp}},
\end{equation}
where \(C_{\parallel}\) is the count rate corresponding to the expected polarization outcome, and \(C_{\perp}\) is that corresponding to the orthogonal outcome within the same basis.
Figs.~\ref{fig3}(a) and \ref{fig3}(b) show the single-photon count rates normalized to the total count rate in each basis.
For the Alice--Charlie link, the measured QBERs are $E_{Z} = (1.68 \pm 0.12)$\% and $E_{X} = (5.62 \pm 0.05)$\%.
For the Bob--Charlie link, they are $E_{Z} = (2.25 \pm 0.25)$\% and $E_{X} = (2.75 \pm 0.06)$\%.
The slight elevation of $E_X$ in the Alice--Charlie link is attributed to minor beam splitter birefringence, which introduces different phase shifts to the $H$ and $V$ components upon reflection and transmission.

Returning the internal HWP to $0^{\circ}$ engages the MDI-QKD operation mode.
Here, data for each encoded polarization state combination were again accumulated over 5~min.
For the MDI-QKD measurements, we denote $C_{ij}^{\pm}$ as the coincidence count for the $\ket{\psi^{\pm}}$ outcome when Alice and Bob encode states $i$ and $j$, respectively, such that total coincidences are $C_{ij} = C_{ij}^{+} + C_{ij}^{-}$.
Based on these definitions, the QBERs for the Z and X bases, $E_{Z}$ and $E_{X}$, are defined as
\begin{align}\label{eq:qber_z}
  E_{Z} = \frac{C_{HH} + C_{VV}}{C_{HH} + C_{VV} + C_{HV} + C_{VH}},
\end{align}
and
\begin{align}\label{eq:qber_x}
  E_{X} = \frac{C_{DD}^{-} + C_{AA}^{-} + C_{DA}^{+} + C_{AD}^{+}}{C_{DD} + C_{AA} + C_{DA} + C_{AD}}.
\end{align}
Fig.~\ref{fig3}(c) shows the measured coincidence count rates normalized by the square of the total detector count rate.
We obtained QBERs of $E_{Z} = (3.5 \pm 1.2)$\% and $E_{X} = (27.5 \pm 0.5)$\% between Alice and Bob.
These results demonstrate reasonable agreement with the corresponding theoretical expectations of 0\% and 25\%, respectively.

To evaluate the performance of our QKD implementation, we calculated the asymptotic secret key rate per sifted signal using the Devetak--Winter formula. \cite{Devetak2005} As previously mentioned, our setup is designed to demonstrate the feasibility of switching between MDI-QKD and BB84 among three parties; thus, we do not employ the decoy-state method in this security analysis. Under the single-photon assumption, the secret key rate is lower-bounded by:
\begin{align}
	r\geq 1-H(E_{Z})-H(E_{X}),
\end{align}
where $H(x)=-x \log_{2}x -(1-x)\log_{2}(1-x)$ is the binary Shannon entropy. The calculated secret key rate per sifted signal range from $0.556$ to $0.574$ for the Alice-Charlie link, $0.647$ to $0.680$ for Bob--Charlie link, and $0$ to $0.000567$ for the Alice--Bob link, representing the best and worst cases considering the experimental standard deviation. Note that we directly used the measured $E_{X}$ to calculate the secret key rate of MDI-QKD. However, in a full implementation of MDI-QKD employing decoy-state method \cite{Hwang2003,Ma2005,Zhao2006}, the single-photon phase error rate would be estimated from photon statistics. Consequently, the error rate in the $X$-basis would be reduced to a similar order of magnitude as that in the $Z$-basis, making the secret key rate of MDI-QKD comparable to that of BB84.

\section{Conclusion}

We have demonstrated a reconfigurable measurement scheme that supports both MDI-QKD and BB84 with a single laser source.
This setup allows a simple transition between polarization encoded MDI-QKD and BB84 modes by toggling a single HWP within the partial BSM module, leaving the remaining optical framework untouched.
By exploiting EOM-based sideband engineering, two mutually independent WCS channels were prepared from a common CW laser, thereby obviating the need for independent laser sources and an active frequency locking scheme.
The measured HOM visibility approached 0.5, the theoretical upper bound for WCSs, which confirms the indistinguishability of the two prepared channels.
In addition, the QBERs obtained in both modes were in good agreement with the theoretical prediction, validating the feasibility of the proposed scheme.
Combining receiver reconfigurability with single-laser operation that eliminates active frequency locking while retaining independent frequency tunability, the proposed scheme provides a practical foundation for flexible and cost-effective quantum networks.

\begin{acknowledgments}
This work was supported by an Agency for Defense Development grant funded by the Korean Government.
\end{acknowledgments}

\end{document}